\documentstyle[12pt]{article}    
\begin{document}
\begin{center}
   
      {\bf One-Loop Effective Action on Rotational Spacetime: Zeta-Function Regularization and Schwinger Perturbative Expansion}

\vspace{3cm}

                       Wung-Hong Huang\\
                        Department of Physics\\
                         National Cheng Kung University\\
                         Tainan, 70101, Taiwan
\end{center}
\vspace{3cm}

  The zeta-function regularization method is used to evaluate the renormalized effective action for massless conformally coupling scalar field propagating in a closed Friedman spacetime perturbed by a small rotation. To the second order of the rotational parameter in the model spacetime the analytic form of the effective action is obtained with the help of  the Schwinger perturbation formula.   After investigating the time evolution of the rotational parameter we find that the quantum field effect can produce an effect which damps the cosmological rotational in the early universe.

\vspace{3cm}
\begin{flushleft}

Classification Number: 04.60+v;11.10.gh\\
 E-mail: whhwung@mail.ncku.edu.tw\\
Annals of Physics 254 (1997) 69
\end{flushleft}

{\newpage} 

\section  {INTRODUCTION}

 	Quantum field plays important role in the dynamical evolution of the early universe [1].     In the epoch between the Planck time ($t\cong 10^{-43} sec$) and grand-unified-theory time ($t\cong 10^{-34} sec$ ) one can treat the gravitational field as a classical background field and the expected values of some quantum-matter stress tensors are regarded as the source of the generalized Einstein equation.   Using this equation the problem of  back reaction of a quantum field in a curved space can then be studied either numerically or analytically.  The central problem in this approach is evaluating the renormalized stress tensor which may be obtained from the metric variational derivative of the renormalized effective action [1].

   Several years ago, Hartle and Hu [2] adopted the dimensional regularization method to evaluate the one-loop contribution of conformally invariant scalar field to the effective action for the homogeneous cosmological model with small anisotropy in the early universe [3].   In a recent paper [4], we extend the method to the spacetime with small inhomogeneity and show that a quantum field can smooth out the space inhomogeneity the early universe. 

 	In this paper we will adopt the $\zeta$-function regularization method [1,5-7] to evaluate the renormalized effective action for massless conformally coupling scalar field propagating in a closed Friedman spacetime perturbed by a small rotation.   Note that in such a model spacetime, as the function form of the zeroth-order field propagator cannot be conformally related to the usual flat space, the Feynman Green function and the mathematical formulation developed in the flat spacetime become useless [6] and method developed by the Hartle and Hu [2] cannot be used to evaluated the effective action. 

   We will evaluate the effective action to the second order of the rotational parameter in the model spacetime with the help of  the Schwinger perturbation formula [8] and then use it to investigate the evolution equation of the rotational parameter.   we have seen that the quantum field effect can produce an effect that damps the cosmological rotational in the early universe.

   Note that the $\zeta$-function regularization method has been extensively used to evaluate the effective action [1,7] and effective potential (for example, for the model in the static Taub spacetime [9], or in the general spacetime with slow variation [10], or in the rotational Godel spacetime [11].  In a recent paper [12] we have also combined it with the  Schwinger perturbation formula to evaluate the effective action in a curved space.  The calculations in that paper are consistent with those evaluated by  the Hartle and Hu [2] for the model in the Bianchi I universe.   However, as the closed Friedman spacetime cannot be conformally transformed to the flat spacetime, the calculation in this paper reveal the advantage of the $\zeta$-function regularization method.   Note also that the application of the Schwinger formula within the $\zeta$-function regularization has been investigated in [13] for the quantum field in the flat spacetime.

   This paper is organized as follows.   In section II we present the model spacetime and formulate the Schwinger perturbation expansion of the renormalized effective action via $\zeta$-function regularization .   In Section III we evaluated the generalized zetz function in detail.  The analytic 
form of the effective action to the second orders of the rotational parameter  is presented.  The possibility of a quantum field effective arising to damp the cosmological rotational in the early universe is discussed in Section IV.   The last section is devoted to a short summary.

\section  {MODEL SPACETIME, EFFECTIVE ACTION AND PERTURBATIVE EXPANSION}

\subsection  {A Rotating Spacetime}
 	The metric of a rotating spacetime used in this paper is given by

     $$d^2s = g_{\mu\nu}dx^{\mu}dx^{\nu} = - dt^2 + r^2 \left[d\chi^2 + sin^2\chi d\theta^2 + sin^2\chi sin^2\theta d\phi^2 - 2sin^2\chi sin^2\theta e(t) d\varphi dt \right].  \eqno{(2.1)}$$
where $e(t)$ is a small quantity representing the metric rotational parameter which is related to the local gagging of inertial frames [14].  The case of $e(t)=0$ is the close Friedmann-Robertson-Walker universe and $r$ is the principal curvature radius of the associated spacetime.  The existence  of the off-diagonal metric term proportional to $d\phi dt $ indicate the existence of space rotation.    Note that to the second order of the rotational parameter $e(t)$ the curvature scalar becomes

$$R= -{6\over r^2}+[2+2cos(2t) -cos(2\chi-2\theta) +2cos(2\theta) -2cos(2\chi-2\theta).  \eqno{(2.2)}$$
Thus the off-diagonal term cannot be removed just by a coordinate transformation.   More discussion about the spacetime properties of the metric form (2.1) can be found in the reference [14].

\subsection  {Effective Action and $\zeta$ -Function}

We consider the Lagrangian describing a massless scalar field with conformal coupling to the gravitational background

$$S= \int L dx^4= \int \sqrt{-g}[- {1\over2}g^{\mu\nu}\partial_\mu\phi\partial_\nu\phi -{1\over12}R\phi^2] d^4x. \eqno{(2.3)}$$
Th field equation is 

$$H\phi=[-g^{\mu\nu}\bigtriangledown_\mu \bigtriangledown_\nu+{1\over6 }R]\phi = 0, \eqno{(2.4)}$$
where R is the Ricci scalar and the $\bigtriangledown$ denotes the covariant derivative.   In the zeta-function regularization the effective action is evaluated by the formula [1,7]

   $$W = -{i\over2}  ln[Det(H)] = -{i\over2} [\zeta'(0) + \zeta(0) ln(\mu^2)].         \eqno(2.5) $$
Using the proper-time formalism, the $\zeta$ function can be evaluated by the relation [1,7]

$$\zeta(\nu) = [\Gamma(\nu)]^{-1} \int dx^4 [-g(x)]^{1/2} \int_0^{\infty}  ids (is)^{\nu -1}<x\mid e^{-isH}\mid x>,    \eqno(2.6)$$
where the operator H is defined in (2.4) and it is understood that $H \rightarrow H - i\epsilon$, with $\epsilon$ a small positive quantity.   The operator H can be expressed as

$$ H=H_0+H_1^{t \varphi}{d^2\over dt d\varphi}+H_1^{\varphi}{d\over  d\varphi} +H_2^{t}{d\over dt }+H_2^{t t}{d^2\over dt^2}+H_2^{t \chi}{d\over d\chi}+H_2^{\varphi \varphi}{d^2\over d\varphi^2}+H_2^{\theta}{d\over \theta}+H_2^{R}+ O(e(t)^3),  \eqno(2.7)$$
where
$$H_0=-{d^2\over dt^2}+{1\over2}(\triangle^{(3)}-1)$$
$$H_1^{t\varphi}=-2e(t);  ~~ H_1^{\varphi}= - \dot {e}(t)$$
$$H_2^t= r^2 sin^2\chi sin^2\theta \dot e(t) e(t)$$
$$H_2^{tt}= r^2 sin^2\chi sin^2\theta \dot e(t) e(t)$$
$$H_2^{tt} ={1\over 4}sin(2\chi)[cos(\theta) -1]e(t)^2$$
$$H_2^{\varphi \varphi}= -e(t)^2; H_2^{\theta}=sin\chi cos\theta e(t)^2$$
$$H_2^{R}=\left({2\over3}-{1\over3}[cos(2\chi)-1][cos(2\theta)-1] \right).  \eqno(2.8)$$
In (2.8) $\triangle^{(3)}$is the Laplace associated with the spatial metric of the closed Friedmann-Robertson-Walker universe.  The eigenfunction $\phi_N$ and eigenvalue $\lambda_n$ for the massless scalar conformally coupling to the closed Friedmann-Robertson-Walker universe are [1]

$$\phi_N= \prod _{KJ}(\chi) Y_J^M(\theta,\varphi), ~~~ N={\omega,J,K,M}, \eqno{(2.9a)}$$
$$\prod _{KJ}(\chi) = [(K^2-1) ... (K^2-J^2)]^{1/2}\left[\phi\over2sin\chi\right]^{1/2} ~P_{K-1/2}^{-1/2-J}, \eqno{(2.9b)}$$
$$\lambda_N = -\omega^2 +{K^2\over r^2}, \eqno{(2.10)}$$
where $Y_J^M$ are the spherical harmonic function and $P_\alpha^\beta$ are the Legendre polynomials, and 

$$K= 1,2,3,...,\infty$$
$$J= 0,1,2,,...,K-1$$
$$M= -J, -J+1,...,J-1,J$$
In the following calculation we occasionally need to use the relations [15]

$$\sum_J \sum_M\prod_{KJ}(\chi)^2 Y_J^M(\theta,\varphi)Y_J^M(\theta,\varphi)^{*}= \sqrt{2\over\pi} K .~~~~~  (angular~~ addition ~~theorm)\eqno{(2.11)}$$
$$\sum_M M^2 Y_J^M(\theta,\varphi)Y_J^M(\theta,\varphi)^{*}= {J(J+1)(2J+1)\over 8\pi}sin^2\theta. \eqno{(2.12)}$$

\subsection {Schwinger Perturbation Expansion}
 The perturbation expansion of effective action can be based on the Schwinger  perturbation formula [8]

 $$Tr e^{-isH} = Tr [ e^{-isH_0} - is e^{-isH_0} H_1  +{s^2\over2}\int_0^{1} du e^{-is(1-u)H_0}H_1 e^{-isuH_0} H_1 $$
  $$+{is^3\over2}\int_0^{1} du \int_0^{1} dv e^{-is(1-u)H_0}H_1 e^{-isu(1-v)H_0} H_1e^{-isuv)H_0} H_1+ ... ] .                 \eqno(2.13)$$
Thus, in our model spacetime the perturbative expansion of the zeta function can be expressed as

$$ \zeta(\nu)= \zeta_0+\zeta_1^{t \varphi}+\zeta_1^{\varphi} +\zeta_{11}^{t \varphi t \varphi}+\zeta_{11}^{\varphi \varphi}+\zeta_2^{t}+\zeta_2^{t t}+\zeta_2^{t \chi}+\zeta_2^{\varphi \varphi}+\zeta_2^{\theta}+\zeta_2^{R}+ O(e(t)^3),  \eqno(2.14)$$
where $\zeta_ n^S$ denotes the generalized $\zeta$ function calculated from $H_n^S$ and $\zeta_{11}^{ST}$ denotes that calculated from combining the term $H_1^S$ and$H_1^T$.   The explicit form and defined calculations of the  generalized $\zeta$ function are presented in the next section.

\subsection   {A more General Spacetime}
It is know that a more cosmologically interesting spacetime is that described the metric

$$ ds^2 = \tilde g_{\mu\nu}dx^\mu dx^\nu = a(t)^2 \left(- dt^2 + r^2 \left[d\chi^2 + sin^2\chi d\theta^2 + sin^2\chi sin^2\theta d\phi^2 - 2sin^2\chi sin^2\theta e(t) d\varphi dt \right]\right). \eqno(2.15)$$
where $a(t)$ denoted the expanding behavior of the cosmology.  As the above metric is related to the metric in (2.1) just by a conformal transformation
 
  $$g_{\mu\nu} \rightarrow \tilde g_{\mu\nu} = e^{2U(x)} g_{\mu\nu}\eqno(2.16)$$
we can use the formula of conformal transformation of renormalized effective action [12]

$$W [\tilde g_{\mu\nu}]= W[g_{\mu\nu}] +{1\over 16 \pi^2} \hbar  \int dx^4 [-g(x)]^{1/2}[ ~A[U(R_{\mu\nu\lambda\delta} R^{\mu\nu\lambda\delta}- 4 R_{\mu\nu} R^{\mu\nu}+ R^2) + 2 RU,_\mu U,^\mu  $$
$$- 4R_{\mu\nu} U,^\mu U,^\nu  - 4 U,_\mu^{;\mu} U,_\lambda U,^\lambda  - 2 (U,_\lambda U'^\lambda )^2 ] + B [U(R_{\mu\nu\lambda\delta} R^{\mu\nu\lambda\delta}- 2 R_{\mu\nu} R^{\mu\nu}+{1\over 3} R^2) +$$
$$ {2\over 3} R(U,_\mu^\mu + U,_\mu U,^\mu ) -  2U,_\mu U,^\mu - 4 U,_\mu^{;\mu} U,_\lambda U,^\lambda  - 2(U,_\lambda U,^\lambda)^2 ]~ ],          \eqno(2.17)$$
to obtain the renormalized effective action in a spacetime with metric $\tilde g_{\mu\nu} $ once we have calculated it in the spacetime time with  metric $g_{\mu\nu}$.    Note that the formula (2.17) has also appeared in several papers, for example [16] and [17], in which some other applications are discussed.

  In the next section we will evaluate the generalized $\zeta$ function for the spacetime with the metric $g_{\mu\nu}(x)$  by using the formula collected in Section IIB. Then, through the conformal transformation formula (2.17), we can obtain the renormalized effective action in the spacetime with metric $\tilde g_{\mu\nu} $.  After the standard variation procedure we get the evolution equation of the rotational parameter $e(t)$ and then analyze the quantized scalar field on the cosmological rotation.   These are investigated in Section IV.

\section  {CALCULATION OF THE GENEALIZED ZETA FUNCTION	}

   We will begin to evaluate the generalized $\zeta$ function.   The zero-order term in the Schwinger perturbation formula is 

$$\zeta(\nu) = {1\over \Gamma(\nu)} \int dx^4 [-g(x)]^{1/2} \int_0^{\infty}  ids (is)^{\nu -1}<x\mid e^{-isH_0}\mid x>$$
  $$ = {1\over \Gamma(\nu)} \int dx^4 [-g(x)]^{1/2} \int_0^{\infty}  ids (is)^{\nu -1}\sum_N \sum_{\tilde N} <x\mid N><N\mid  e^{-isHH_0}\mid{\tilde N}><{\tilde N}\mid  x>$$
  $$ = {i\over \Gamma(\nu)} \int dx^4 [-g(x)]^{1/2} \int_0^{\infty}  ds (s)^{\nu -1}\int {d\omega\over 2\pi }\sum_K \sqrt{2\over\pi} K exp[-s(\omega^2 +{K^2\over r^2})]$$
  $$ = {i\over 2\pi} \int dx^4 [-g(x)]^{1/2} \sum_K \sqrt{2\over\pi} K \Gamma(\nu -{1\over2})\left({K^2\over r^2}\right)^{1/2-\nu} \Gamma(\nu)^{-1}$$
  $$ = -i\nu  \int dx^4 [-g(x)]^{1/2} {1\over r}\sum_K \sqrt{2\over\pi} K^2 +O(\nu^2)$$
$$= \sim  \zeta_R(-2),   \eqno(3.1)$$
where $\zeta_R(\nu)$ is the Riemann zeta function. Note that to obtain the above result we have carried the integration by letting $s \rightarrow i s$ and rotating $\omega $ through  ${\pi\over2}$ in the complex plane.  Also, we retain only the terms proportional to $\nu$ as the renormalized effective action $W= -i/2  ~[\zeta^{\prime} (0)]$ if we choose $\mu=1$.

   The above procedure can also be used to evaluate other $\zeta$ function and we have 

$$\zeta_2^{\varphi\varphi}(\nu)= [\Gamma(\nu)]^{-1} \int dx^4 [-g(x)]^{1/2} \int_0^{\infty}  ids (is)^{\nu -1}(-is) <x\mid e^{-isH_0}H_2^{\varphi \varphi}{d^2\over d\varphi^2}\mid x>$$
$$= {-i\over 6 \Gamma(\nu)\pi} \int dr r^3 e^2(t) \int_0^{\infty} ds s^{\nu-1} \int d \omega \sum_K\sum_J (2J^3+3J^2+J) e^{-s(\omega^2+{K62\over r^2})}$$
$$=  {-i\nu\over12} \int dr r^4 e^2(t)  \sum_K [K^3-K]$$
$$= {-i\nu\over12} \int dr r^4 e^2(t) \left({1\over12}+{1\over120}\right).    \eqno(3.2)$$
Some trivial results are as follows

$$\zeta_2^{tt}(\nu)= [\Gamma(\nu)]^{-1} \int dx^4 [-g(x)]^{1/2} \int_0^{\infty}  ids (is)^{\nu -1}(-is) <x\mid e^{-isH_0}H_2^{tt}{d^2\over dt^2}\mid x>$$
$$=\sim \zeta_R(-2) =0.$$

$$\zeta_1^{\varphi\varphi}(\nu)= [\Gamma(\nu)]^{-1} \int dx^4 [-g(x)]^{1/2} \int_0^{\infty}  ids (is)^{\nu -1}(-is) <x\mid e^{-isH_0}H_2^{\varphi }{d\over d\varphi}\mid x>$$
$$= \sim _M Y_J^M(\theta,\varphi)Y_J^M(\theta,\varphi)^* =0.$$

$$\zeta_2^{\chi}(\nu)= [\Gamma(\nu)]^{-1} \int dx^4 [-g(x)]^{1/2} \int_0^{\infty}  ids (is)^{\nu -1}(-is) <x\mid e^{-isH_0}H_2^{\chi}{d\over d\chi}\mid x>$$
$$=\sim {1\over d \chi}\left[\sum_J \sum_M\prod_{KJ}(\chi)^2 Y_J^M(\theta,\varphi)Y_J^M(\theta,\varphi)^*\right]=0.$$

$$\zeta_2^{\theta}(\nu)= [\Gamma(\nu)]^{-1} \int dx^4 [-g(x)]^{1/2} \int_0^{\infty}  ids (is)^{\nu -1}(-is) <x\mid e^{-isH_0}H_2^{\theta}{d\over d\theta}\mid x>$$
$$=\sim {1\over d \theta}\left[\sum_J \sum_M\prod_{KJ}(\chi)^2 Y_J^M(\theta,\varphi)Y_J^M(\theta,\varphi)^*\right]=0.$$

$$\zeta_2^{R}(\nu)= [\Gamma(\nu)]^{-1} \int dx^4 [-g(x)]^{1/2} \int_0^{\infty}  ids (is)^{\nu -1}(-is) <x\mid e^{-isH_0}H_2^{R}\mid x>$$
$$=\sim \int d \chi \int d\theta  sin^2\chi sin\theta [2-(cos2\chi-1)(cos2\theta-1)]=0.$$
It can easily be seen that 

$$\zeta_1^{t\varphi}(\nu)= [\Gamma(\nu)]^{-1} \int dx^4 [-g(x)]^{1/2} \int_0^{\infty}  ids (is)^{\nu -1}(-is) <x\mid e^{-isH_0}H_1^{t\varphi }{d^2\over dt d\varphi}\mid x>=0,$$

$$\zeta_2^t(\nu)= [\Gamma(\nu)]^{-1} \int dx^4 [-g(x)]^{1/2} \int_0^{\infty}  ids (is)^{\nu -1}(-is) <x\mid e^{-isH_0}H_2^t{d\over d t}\mid x>=0,$$
as these function contain the derivative $d/dt$ which will produce a factor $\omega$,  and thus we have an integration over an odd function of variable $\omega$.

 	We next tern to the remain two generalized $\zeta$ function, which involving more calculations.  

$$\zeta_{11}^{\varphi\varphi}(\nu)= \Gamma(\nu)^{-1} \int dx^4 [-g(x)]^{1/2} \int ids (is)^{\nu -1}(-{s^2\over 2})\int_0^1 du  <x\mid e^{-is(1-u)H_0}H_1^\varphi \partial\varphi e^{-isuH_0}H_1^\varphi \partial\varphi \mid x>$$
$$= {1\over\Gamma(\nu)}\int  dt~ r^3 \int d\tilde t \dot e(t) \dot e(\tilde t) \int {d\omega\over2\pi}\int {d\tilde\omega\over2\pi} exp[-i(\omega-\tilde\omega)] \sum_K (K^4-K^2) \int ds (s)^{\nu +1}\int_0^1 du$$
$$ exp[-s(1-u)(\omega^2+{K^2\over r^2})] exp[-su(\tilde \omega^2+{K^2\over r^2})] $$
$$= {-\nu\over12} \int  dt~ r^3 \int d\tilde t \dot e(t) \dot e(\tilde t) \sum_K (K^4-K^2) \int {d\omega\over2\pi}\int {d\tilde\omega\over2\pi} {1\over {\omega^2-\tilde\omega^2}}$$
$$\left[(\omega^2+{K^2\over r^2})^{-\nu-1} - (\tilde \omega^2+{K^2\over r^2}){-\nu-1}\right] exp[-i(\omega -\tilde\omega )(t-\tilde t)] $$
To obtain the above result we have first integrate $u$ then shift $\tilde\omega\rightarrow \omega+\tilde\omega$ and then integrate the variable $s$.

  As the above relation cannot not be integrate exactly, we will evaluate it separately for the small$r$ and large $r$ region.

  (i)   For the case of small $r$, which represents a universe with a small radius, we can use the formula derived in Appendix A-1 to integrate the variable $\tilde\omega$ .  Then by integrate the variable $\omega$ and taking the summation of K, we have the result

$$\zeta_{11}^{\varphi\varphi}(\nu)= {i\over 96} \left( 1+\nu\left({17\over 6} -2ln2+2\gamma +2 lnr \right)  \right) \int dt r^6 \dot e(t)^2, +O(\nu^2),   \eqno{(3.3)}$$
where $\gamma$ id the Euler constant.

  (ii)   For the case of large $r$, which represents a universe with a large radius, we can regard $K/r$ as a continuum variable and integrate the variable $K$ first, then the variable $\tilde\omega$, and finally the variable $\omega$.   We therefore have the result

$$\zeta_{11}^{\varphi\varphi}(\nu)= {i\over 192} \left( 1+\nu\left({8\over 3} +\pi ~i  \right)  \int dt~\dot e(t)^2 -{\nu\over 2\pi} \int dt dr ~d \tilde t d\omega  \ddot e(t)\ddot e(\tilde t)~cos(\omega(t-\tilde t))\right)$$
$$ - i{\pi r^5\over 482} \left( (1+\nu (2+\pi i)) \int dt~\dot e(t)^2 -{\nu\over 2\pi} \int dt dr d \tilde t d\omega ~ \dot e(t)\dot e(\tilde t)~cos(\omega t))  ln\mid \omega^2\mid\right).  \eqno{(3.4)}$$

Note that in the case of small $r$ the $\zeta$ function calculated in the above are pure imaginary and thus the effective action does not have an imaginary part.   Therefore it does not have particle production in this epoch [1].   This is because that constraining the system in the small $r$ region is equivalent to restrain it in the states close to the vacuum, and the universe has not evolve far from vacuum thus the particle has not yet been produced.   On the other hand, for the case of large $r$ region the universe has evolve far from the in-vacuum state and the effective action does have  an imaginary part, as the calculation above show,  and thus particles are produced.

The same procedure can be used to obtain the final $\zeta$ function

$$\zeta_{11}^{t \varphi t \varphi}(\nu)= \Gamma(\nu)^{-1} \int dx^4 [-g(x)]^{1/2} \int ids (is)^{\nu -1}(-{s^2\over 2})\int_0^1 du  <x\mid e^{-is(1-u)H_0}H_1^{t \varphi}\partial_ t \partial_\varphi e^{-isuH_0}H_1{t \varphi}\partial_t \partial_\varphi  \mid x>$$
$$= {\nu\over 3} \int  dt~ r^3 \int d\tilde t \dot e(t) \dot e(\tilde t) \sum_K (K^4-K^2) \int {d\omega\over2\pi}\int {d\tilde\omega\over2\pi} {1\over {\omega^2-\tilde\omega^2}}$$
$$\left[(\omega^2+{K^2\over r^2})^{-\nu-1} - (\tilde \omega^2+{K^2\over r^2}){-\nu-1}\right] exp[-i(\omega -\tilde\omega )(t-\tilde t)] $$
To proceed, we will evaluate the above relation separately for the small $r$ and large $r$ region.

    (i)   For the case of small $r$, we shift $\tilde\omega \rightarrow \omega +\tilde\omega$ and  use the formula derived in Appendix A-2 to integrate the variable $\tilde\omega$ .  Then after the integration of  the variable $\omega$ and summation of K, the above relation becomes

$$\zeta_{11}^{t \varphi t \varphi}(\nu)= {i\over 96} \left( 1+\nu\left({17\over 6} -2ln2+2\gamma +2 lnr \right)  \right) \int dt r^6 \dot e(t)^2 - {i  11\over 11520} \nu \int dt r^4  e(t)^2 +O(\nu^2),   \eqno{(3.5)}$$

  (ii)   For the case of large $r$  we can regard $K/r$ as a continuum variable and integrate the variable $K$ first, then the variable $\tilde\omega$, and finally the variable $\omega$.   We find that in this case the vale of $\zeta_{11}^{t \varphi t \varphi}(\nu)$ is equal to that of $\zeta_{11}^{\varphi \varphi}(\nu)$ expressed in (3.4).    This completes our calculations  about the generalized zeta function.

   Finally, we collect the above calculation of $zeta$ function and substitute them into the formula  (2.5).  Thus we obtain the renormalized effective action $W[g_{\mu\nu}]$.

   (i)   For the case of small $r$.

  $$W[g_{\mu\nu}] = \int dt \left(- {99\over 23040}r^4~ d(t)^2 + {1\over 96}  \left({17\over6} -2ln2 + 2\gamma +2lnr\right) r^6 \ddot e(t)^2\right).   \eqno(3.6)$$

 (i)   For the case of large $r$.

  $$W[g_{\mu\nu}] = \int dt \left(- {99\over 23040}r^4~ d(t)^2 + {1\over 96}  \left({17\over6} -2ln2 + 2\gamma +2lnr\right) r^6 \ddot e(t)^2\right)$$
$$+\left( {\pi r^7\over 384} ({8\over 3} +i \pi) - {\pi r^5\over 964} (2 +i \pi)\right ) \int dt~\dot e(t)^2 - {1\over 2\pi} \left({\pi r^7\over384}- {\pi r^5\over 964}\right) \int dt dr d \tilde t d\omega ~ \dot e(t)\dot e(\tilde t)~cos(\omega t))  ln\mid \omega^2\mid. \eqno(3.7)$$

\section  {DAMPING OF THE COSMOLOGICAL ROTATION}

    We now use the formula (2.170 to find that the renormalized effective in the spacetime with metric $\tilde g_{\mu\nu}$, described in  (2.15) is 

  $$\tilde W[\tilde g_{\mu\nu}] = W[g_{\mu\nu}]  + {1\over1920}\int dt {3\over 4}r^2\left({\dot a^2 \ddot a\over a^3}- 3{\ddot a\over a^3}\right) + r^5 \left(3 {\dot a^2 \over a^2}-6 {\dot a^2 \ddot a\over a^3}+5{\dot a^4 \over a^4}-({\dot a^2 \over a^2})\ddot{}\right) e^2 + Oe^3).   \eqno{(4.1)}$$
Substituting  the results of  $W[g_{\mu\nu}]$derived in (3.6) and adding the classical action in to   $W[\tilde g_{\mu\nu}]$, the final expression of the total effective action is thus obtained.  The evolution equation of the rotating parameter can be obtained by varying the total effective action with respect to the function $e(t)$.   We separately discuss the equation for the large $r$ region and small $r$ region in the following.

 \subsection {Small $r$ Region}
For the case of small $r$  the evolution equation is

 $$ m \ddot e(t) + k(t) d(t) = 0.\eqno{(4.2)}$$
with 
 $$ m = {1\over 48}  \left({17\over6} -2ln2 + 2\gamma +2lnr\right) r^6 .\eqno{(4.3)}$$
 $$ k(t) = {11\over1280}r^4 - {1\over960} r^5 \left(3 {\dot a^2 \over a^2}-6 {\dot a^2 \ddot a\over a^3}+5{\dot a^4 \over a^4}-({\dot a^2 \over a^2})\ddot{}\right) +{3\pi \over 8G} (r^3 a^2 + r^5 \dot a^2), \eqno{(4.4)}$$
where G is the gravitational constant.   The qualitative property of the evolution of a cosmological rotation parameter $e(t)$ can be determined with the following analysis.  

   First, we know that although the value of $r$ is small in the early universe, it is not a very small number, as we are interesting in the epoch after the Planck time.   Thus the value of $m$ in (4.3) is positive.  Next, if we adopt $a(t) =t$, which represents the cosmological expansion by classical radiation then $k(t)$ will be increasing during th cosmological evolution.  (Note that, in general, we include classical radiation to support the expansion of the universe at times late compared with the Planck time.)  We can now regard (4.1) as a harmonic oscillator with positive mass $m$ and an increasing string strength $k(t)$.   Then, because there is no external driving force in this string, the total energy of the string will be conservative.  Thus, the rotating parameter $e(t)$ shall behave as an oscillator with decreasing amplitude.

 \subsection {Large  $r$ Region}
For the case of  lage  $r$  the evolution equation is

 $${\pi r^7\over48} ({8\over3}+\pi i ){\dot{}} \ddot e {\dot{}} -  {\pi r^5\over12} (2+\pi i )\ddot e  +\left( -{11\over1440}r^4  +  {r^5\over960} \left(- {\dot a^4 \over a^4}+4 {\dot a^2 \ddot a\over a^3}+{\ddot a^3 \over a^2}-2 {\dot a \ddot a {\dot{}} \over a^2}\right) -{3\pi \over 8G} r^5 a^2 \right)~ e  =0.   \eqno{(4.5)}$$
To obtain the above equation we have neglected the nonlocal terms.   This approximation is hard to justify, but  the  Hartle [2] paper has shown that it does not change the results significantly.  For the case of $a(t) = t$ and large $r$ the above equation becomes 

 $$ r^2 ({8\over3}+\pi i ){\dot{}} \ddot e {\dot{}} -  (8 + 4\pi i) \ddot e - {18 \over G} e \cong 0.   \eqno{(4.6)}$$
As the above equation is a linear differential equation with constant coefficients it can be solved exactly.   It is easy to see that the solution is

$$ e(t) = c_1~sin(a_1 t)e^{b_1t}+c_2~sin(a_2 t)e^{b_2t}+c_3~sin(a_3 t)e^{b_3t}+c_3~sin(a_3 t)e^{b_3t},  \eqno{(4.7)}$$
where $c_i$ is the integration constant, and $a_i$ and $b_i$ are positive constants which can be easily determined by (4.6) but are too complex to be cited here.

  Equation (4.7) tells that there are solutions with $c_1=c_3=0$, in which the rotating parameter $e(t)$ behave as an oscillator with decreasing amplitude.  The other solutions which behave as exponentially growing oscillator solutions, i.e., $c_1\ne 0$  and/or $c_3\ne 0$, are beyond the scope of the perturbation approach used in this paper and shall be neglected.   We thus see that the quantum field can produce an effect which damps the cosmological rotation of the early universe.

\section  {CONCLUSION}
   In this paper we have applied the zeta-function regularization method to evaluate the renormalized effective action for massless conformally coupling scalar field propagating in a closed Friedman spacetime perturbed by a small rotation.   With the help of the Schwinger perturbation formula we obtain the analytic form of the effective action to the second order of the rotational parameter in the model spacetime.  Due to the mathematical difficulty, the closed form of the effective action can only be found for the case of the small-limit value (i.e., near the in-vacuum epoch) and large-limit (i.e., far from the in-vacuum epoch) of the principal curvature radius of the spacetime.

    We also investigate the time evolution behavior of the rotational parameter.  It is found that, for that case of a small principal curvature radius, the effective action is a real function and no particles are produced.   The rotation parameter $e(t)$ behaves as an oscillator with decreasing amplitude.   However, constraining the system to near vacuum states does not have much cosmological interest and we next consider the case far from vacuum.  For the case of large principal curvature radius, we see that the effective action is a complex function and there are particle production.   The consistent solutions shows that the rotating parameter will behave as an exponentially damping oscillator and thus the quantum field can  produce an effect which damps out the cosmological rotation in the early universe.
%%%%%%%%%%%%%%%

%%%%%%%%%%%%%%%%%%%%%%%%
\newpage
\begin{center} {APPENDIX:  TWO INTEGRATION FORMULAS}
\end{center}
$$\int {d\tilde\omega\over 2\pi}\left(\tilde\omega^2 - {\omega^2\over 4}\right)^{-1}\left(\tilde\omega^2 +{K^2\over 4r^2}\right)^{-\nu-1} $$
$$= -(2r)^{3+2\nu} K^{-3-2\nu}\pi^{-1/2}\Gamma(\nu +{3\over2}) [\Gamma(1+\nu)]^{-1}~ {{}_2}F_1\left(1,\nu+{2\over3},{2\over3},-{r^2\omega^2\over K^2} \right)$$
$$= \left(-{4r^3\over  K^{3+2\nu}}+{4\omega^2 r^5\over  K^{5+2\nu}}\right) \left(1+\nu ({8\over3} + 2 ln r)) \right) + O(\nu^2) +O(r^6).  \eqno{(A.1)}$$
In the above equation ${{}_2}F_1$ is the hypergeometric function [17].

$$\int {d\omega\over 2\pi}\int {d\tilde\omega\over 2\pi}  {\tilde\omega(\omega+\tilde\omega)\over  \tilde\omega(\omega+2 \tilde\omega)} \left(\tilde\omega^2 +{K^2\over 4r^2}\right)^{-\nu-1} exp[\omega(t-\tilde t)] $$
$$=\int {d\omega\over 2\pi}\int {d\tilde\omega\over 2\pi}  {\tilde\omega(\omega+\tilde\omega)(\omega-2\tilde\omega)\over  \tilde\omega(\omega+2 \tilde\omega)(\omega-2\tilde\omega)} \left(\tilde\omega^2 +{K^2\over 4r^2}\right)^{-\nu-1} exp[\omega(t-\tilde t)] $$
$$=\int {d\omega\over 2\pi}\int {d\tilde\omega\over 2\pi} \left( {-\tilde\omega^2 \over \omega^2-4 \tilde\omega^2}+ odd~function ~of ~ \tilde\omega\right) \left(\tilde\omega^2 +{K^2\over 4r^2}\right)^{-\nu-1} exp[\omega(t-\tilde t)] $$
$$=  \int {d\omega\over 2\pi}\int {d\tilde\omega\over 2\pi} {1\over4} \left(\tilde\omega^2 +{K^2\over 4r^2}\right)^{-\nu-1} exp[\omega(t-\tilde t)]+\int {d\omega\over 2\pi}\int {d\tilde\omega\over 2\pi} {\omega^2\over16} \left(\tilde\omega^2 -{\omega^2\over 4}\right)^{-1}\left(\tilde\omega^2 +{K^2\over 4r^2}\right)^{-\nu-1} exp[\omega(t-\tilde t)] $$
$$=i r {1\over 8K}\delta(t-\tilde t) + i \int {d\omega\over 2\pi}{\omega^2\over 16}\left( -{4r^3\over K^{3+2\nu}} -{4\omega^2 r^5\over K^{5+2\nu}} \right) \left(1+\nu({8\over3} +2 ln r)  \right)  exp[\omega(t-\tilde t)].\eqno{(A.2)}$$
The final result is obtained by using the formula (A1) and rotating $\omega$ back through an angle of $-\pi/2$ in the complex plane.
\newpage

\begin{enumerate}

\item  N. D. Birrell and P. C. W. Davies, Quantum Field in Curved Space (Cambridge University Press, Cambridge, England, 1982)
\item  J. B. Hartle, Phys. Rev. Lett. 39, 1373 (1977); M.V. Fischetti. J. B. Hartle and B. L. Hu, Phys. Rev. D 20, 1757 (1979); ;J. B. Hartle and B. L. Hu, Phys. Rev. D 20, 1772 (1979); D21, 2756 (1980).
\item   J. B. Hartle, Phys. Rev. D 22, 2091 (1980)
\item   W. H. Huang, Phys. Rev. D 48, 3914 (1993). 
\item   S. Hawking, Commum. Math. Phys. 55, 133 (1976)
\item   P. Ramond, "Field Theory", Addison Wesley, Redwood City. CA 1989.
\item    L. Parker, " Aspects of quantum filed theory in curved spacetime: effective action and energy-momentum tensor" in "Recent Developments in Gravitation", ed. S. Deser and M. Levey (New York: Plenum, 1977).
\item    J. Schwinger, Phys. Rev. 82, 664 (1951).
\item  T. S. Shen, B. L. Hu, and D. J. O'Connor, Phys. Rev. D 31, 2401 (1985); A. Stylianopoulos, Phys. Rev. D 40, 3319 (1989).
\item    A. L. Berkin, Phys. Rev. D 46, 1551 (1992); K. Kirstein, G. Cognola, and L. Vanzo,   Phys. Rev. D 48, 2813 (1993);  W. H. Huang, Class. Quantum Grav. 10, 2021 (1993);G. Cognola, Phys. Rev. D 50, 909 (1994)
\item  W. H. Huang, Class. Quantum Grav. 8, 1471 (1991)
\item    W. H. Huang, Conformal Transformation of Rnormalized Effective in Curved Spacetimes, Phys. Rev. D 51, 579 (1995). 
\item D. G. C. McKeon and T. N.  Sherry, Phys. Rev. D 35, 3854 (1987).
\item S. S. Bayin and F. L. Cooperstock,  Phys. Rev. D 22, 2317 (1980).
\item V. Fock, Z. Phys. 98, 148 (1935); D. A. Varshalovich, A. N. Moskalev, and V. K. Khersonskii, "Quantum Theory of Angula Momentum", World Scientific, Singapore, 1988.
\item  M. R. Brown and  A. C. Ottewill,  Phys. Rev. D 31, 2514 (1985)

\item  J. S. Dowker, Phys. Rev. D 33, 3150 (1986); J. S. Dowker nad L. P. Schofield, , Phys. Rev. D 38, 3327 (1988)
\item    I. S. Gradshteyn and I. M. Ryzhik ,"Table of Intergals, Series and Products." Academic Press. New York 1980.

\end{enumerate}
\end{document}